\documentclass[]{article}  %

\usepackage{arxiv_style}

\usepackage[utf8]{inputenc} 
\usepackage[T1]{fontenc}    
\usepackage{hyperref}       
\usepackage{url}            
\usepackage{amsfonts}       
\usepackage{nicefrac}       
\usepackage{microtype}      
\usepackage{doi}
\usepackage[numbers,sort&compress,merge]{natbib}
\usepackage{caption}
\usepackage{jabbrv}
\usepackage{amsmath}
\usepackage{amssymb}
\usepackage{graphicx}
\usepackage{siunitx}
\usepackage{url}
\usepackage{xfrac}
\DeclareMathOperator{\arctantwo}{arctan2}

\newcommand{\T}[1]{\textup{#1}}
\newcommand{\NphiSphi}[2]{\ensuremath{\T{N}_{#1}\T{S}_{#2}}}
\newcommand{\diff}[1]{\ensuremath{\T{d}#1}}
\newcommand{\ddiff}[2]{\ensuremath{\T{d}#1\,\T{d}#2}}
\newcommand{\ipd}{\ensuremath{\Delta\varphi}}
\newcommand{\ild}{\ensuremath{\Delta l}}

\graphicspath{{./graphics/}}

\title{
Statistics of the interaural parameters for dichotic tones in diotic noise
(\NphiSphi{0}{\psi})}

\author{ \href{https://orcid.org/0000-0002-6067-1602}{\includegraphics[scale=0.06]{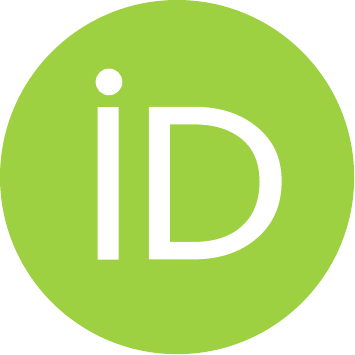}\hspace{1mm}Jörg.~Encke} \\
Department für Medizinische Physik und Akustik\\
Universität Oldenburg\\
26111 Oldenburg, Germany\\
	\texttt{joerg.encke@uni-oldenburg.de} \\
	\And
	\href{https://orcid.org/0000-0002-1830-469X}{\includegraphics[scale=0.06]{orcid.pdf}\hspace{1mm}Mathias.~Dietz} \\
Department für Medizinische Physik und Akustik\\
Universität Oldenburg\\
26111 Oldenburg, Germany\\
}

\begin{document}
\maketitle

\begin{abstract}
  Stimuli consisting of an interaurally phase-shifted tone in diotic
  noise - often referred to as $\NphiSphi{0}{\psi}$ - are commonly
  used in the field of binaural hearing. As a consequence of mixing
  diotic noise with a dichotic tone, this type of stimulus contains
  random fluctuations in both interaural phase- and level-difference.
  This study reports the joint probability density functions of the
  two interaural differences as a function of amplitude and interaural
  phase of the tone. Furthermore, a second joint probability density
  function for interaural phase differences and the instantaneous
  power of the stimulus is derived.
\end{abstract}

\maketitle

\section{Introduction}
Tone in noise detection thresholds improve when the interaural
configuration of tone and noise differ compared to the diotic case. A
rich literature reports on the influence of virtually every parameter
of acoustic stimuli on this binaural unmasking \cite[see e.g.][for a
review]{Culling2021}. Amongst these parameters the phase difference
$\psi$ introduced between the target tones of the two ear-signals is
fundamental and was explored already in the first study of dichotic
tone in noise detection by \citet{Hirsh1948}. Such a signal is commonly
referred to as \NphiSphi{0}{\psi} where the subscripts indicate the
interaural phase difference (IPD) of the noise (N) or signal (S). The
difference between the detection threshold for the purely diotic
\NphiSphi{0}{0} and the \NphiSphi{0}{\psi} signal is referred to as
the binaural masking level difference (BMLD) and is largest for the
case where $\psi=\pi$ \cite{Hirsh1948}.

Adding a dichotic $S_{\psi}$ tone to diotic $N_{0}$ noise introduces
an incoherence between the left and right signals -- which in turn
results in random fluctuations of the interaural phase and level
differences (IPD, ILD) (visualized in
Fig.~\ref{fig:schematic}(a)). The incoherence increases with the tone
level so that binaural unmasking and incoherence detection are often
treated synonymously \cite{Durlach1986}. The value of interaural coherence
itself however was found to be an insufficient predictor for
incoherence detection performance. Instead, detection performance
correlated with the amount of IPD and ILD fluctuations as measured by
the standard deviation \cite{Goupell2006}. Knowledge about the
statistical processes that underlay these fluctuations can thus be
instrumental for binaural modeling as well as stimulus design. Only
relatively few studies, however, have previously treated these
statistics. The probability density function (PDF) underlying the
statistical distribution of IPDs in incoherent noise has been derived
in the frame of optical interferometry
\cite{Just1994}. \citet{Henning1973} derived the PDF for IPDs in the
special case of \NphiSphi{0}{\pi} and using a very similar approach
for the same stimulus condition, \citet{Zurek1991} additionally
derived marginal PDFs for ILDs. Other studies seemed to also have
worked on stimuli where the tone IPD did not equal $\pi$ but his work
seemed to have remained unpublished \cite{Levitt1966}. This study
closes this gap by deriving a closed form expression for the joint PDF
of IPDs and ILDs in the general case of a $N_{0}S_{\psi}$
stimulus. From this distribution, the marginal PDFs can also be
calculated by means of numerical integration.

If fluctuations of the IPD are indeed a cue used to detect the tone in
an \NphiSphi{0}{\psi} stimulus, then the stimulus energy at which
these fluctuation occurred might also affect performance. The product
of the left and right ear stimulus envelope, here called $P'$, can be
interpreted as a measure of instantaneous stimulus
power. Consequently, this study also drives the joint PDF for $P'$ and
IPD.

\section{Deriving the probability density functions}

\begin{figure*}[t]
  \centering
  \centerline{\includegraphics{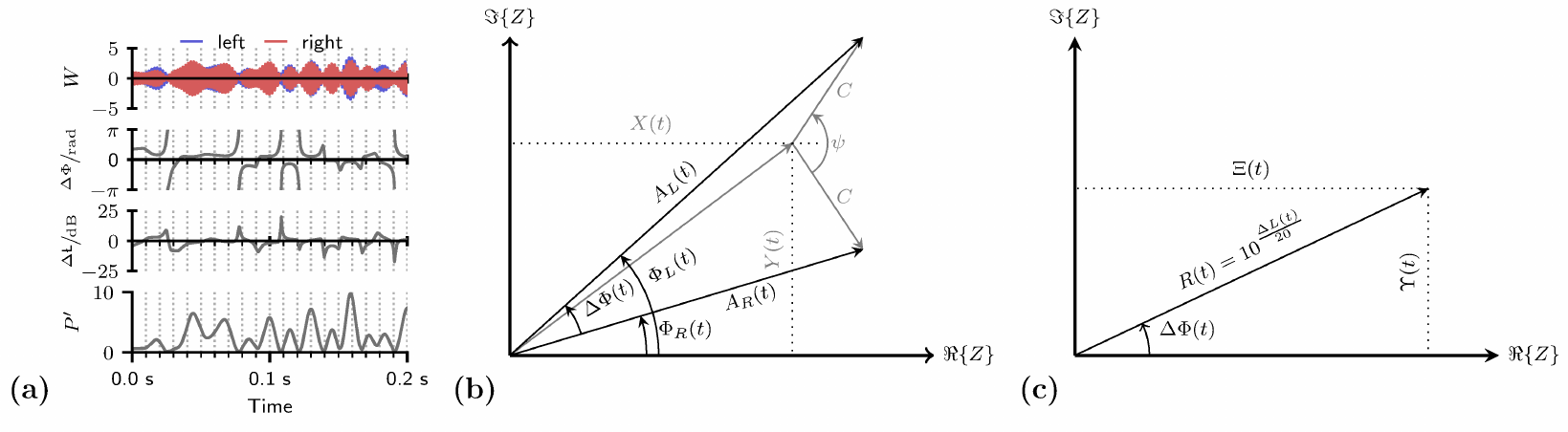}}
  \caption{\textbf{(a)} Visualization of the random fluctuations in
    IPD $\Delta\Phi(t)$ and ILD $\Delta L(t)$  and $P'$ due to mixing an antiphasic
    500Hz tone with a \SI{500}{\hertz} wide band of diotic noise
    (SNR=\SI{-10}{\decibel}). \textbf{(b)} Complex-plane representation of
    the components in a \NphiSphi{0}{\psi} stimulus. The
    left-ear-signal is constructed by adding a ``tone''-vector with
    length $C$ and angle $+\sfrac{\psi}{2}$ to the noise vector
    $X(t)+iY(t)$. The right-ear-signal is constructed by adding a
    ``tone''-vector with an angle of $-\sfrac{\psi}{2}$ to the same
    noise. \textbf{(c)} Complex-plane representation of the
    interaural-baseband $Z(t)=\Xi(t)+i\Upsilon(t)$ which is gained by
    dividing the left-ears-baseband by the right-ears-baseband. The
    absolute value of the baseband equals the interaural amplitude
    ratio $R$ while the phase equals the interaural phase difference
    $\Delta\Phi$.}
\label{fig:schematic}
\end{figure*}

If $N$ is a Gaussian bandpass noise process with a mean value of zero,
the process can be represented using its in-phase and quadrature
components $X$ and $Y$:
\begin{align}
  N(t) = X(t)\cos(\omega_0 t) - Y(t)sin(\omega_0 t),
\end{align}
where $X$ and $Y$ are orthogonal noise processes with the same
variance and mean as $N$. The reference frequency $\omega_{0}$ is
conveniently chosen to equal the frequency of the tone
$S(t)=C\cos(\omega_{0}t+\psi)$ which is added with the amplitude $C$
and phase $\psi$. The resulting stimulus can then be expressed as:
\begin{align}
  W(t) =& \left[X(t)+C\cos(\psi)\right]\cos(\omega_0 t) \nonumber\\
  &- [Y(t)+C\sin(\psi)]\sin(\omega_0 t). \label{eq:w_t}
\end{align}
Or alternatively, employing the signals complex baseband $Z(t)$:
\begin{align}
  W(t) &= \Re\left\{Z(t)\T{e}^{i\omega_0 t}\right\}\\
  Z(t) &= \left[X(t)+C\cos(\psi)\right]
         + i \left[Y(t)+C\sin(\psi)\right] \nonumber\\
  &= A(t) \T{e}^{i\Phi(t)}\label{eq:cplx_env}
\end{align}
where $\Re(x)$ extracts the real part, $i=\sqrt{-1}$ is the imaginary
unit and $A(t)$, $\Phi(t)$ are the instantaneous amplitude and
phase of the baseband.

In case of the \NphiSphi{0}{\psi} stimulus, a cosine with phase
$\sfrac{\psi}{2}$ is added to the noise in the left-ear signal while
the phase of the cosine in the right-ear signal is
$\sfrac{-\psi}{2}$. This process is shown in
Fig.~\ref{fig:schematic}(b) where the individual components are
visualized as vectors in the complex plane. Applying
Eq.~\eqref{eq:cplx_env} results in the complex basebands for the left-
and right-ear signals $Z_{L}(t)$ and $Z_{R}(t)$.

Based on these expressions, PDFs for the interaural parameters will be
derived using two separate approaches. In the first approach, the
baseband of the left-ear signal $Z_L(t)$ is divided by the baseband of
the right-ear signal $Z_R(t)$ resulting in the interaural baseband
$Z_{1}(t)$:
\begin{align}
  Z_{1}(t) &= \frac{Z_R(t)}{Z_L(t)} = \frac{A_R(t)}{A_L(t)}\T{e}^{i[\Phi_R(t)-\Phi_L(t)]}\nonumber\\
  &=R(t) \T{e}^{i\Delta\Phi(t)},\label{eq:ia_eq}
\end{align}
where $\Delta\Phi(t)$ and $R(t)$ are the instantaneous IPDs and
the interaural amplitudes ratios (IARs) respectively. Instantaneous
ILDs can then be calculated as: $\Delta L(t) =20 \log_{10}R(t)$.
To derive the PDF for IPDs and the product of the left and right-ear
envelope $P'$, $Z_{L}(t)$ is multiplied with the complex conjugate of
$Z_{R}(t)$ resulting in
\begin{align}
  Z_{2}(t) &= Z_R(t)Z^{*}_L(t) = A_R(t)A_L(t)\T{e}^{i[\Phi_R(t)-\Phi_L(t)]}\nonumber\\
  &=P'(t) \T{e}^{i\Delta\Phi(t)},\label{eq:ia_eq2}
\end{align}
The process of deriving the PDFs from Eq.~\eqref{eq:ia_eq} and
Eq.~\eqref{eq:ia_eq2} follows the exact same rational so that the
process will only be detailed for Eq.~\eqref{eq:ia_eq} the results
for the second approach will then be stated without further detail.

For the interaural baseband, $Z_{L}$ and $Z_{R}$ as resulting form
Eq.~\eqref{eq:w_t} are inserted into Eq.~\eqref{eq:ia_eq} resulting
in:
\begin{align}
  Z_{1}(t) &=  \Xi(t) + i\Upsilon(t)\nonumber\\
           &= \frac{C \cos{\left(\sfrac{\psi}{2} \right)}
             + i \left[C \sin{\left(\sfrac{\psi}{2} \right)}
             + Y{\left(t \right)}\right]
             + X{\left(t \right)}}
             {C \cos{\left(\sfrac{\psi}{2} \right)}
             - i \left[C \sin{\left(\sfrac{\psi}{2} \right)}
             - Y{\left(t \right)}\right]
             + X{\left(t \right)}} \label{eq:ia_baseband}
\end{align}
\begin{align}
  \Xi(t) &= \frac{Y^{2}{\left(t \right)}
           + \left[C \cos{\left(\sfrac{\psi}{2} \right)}
           + X{\left(t \right)}\right]^{2}
           - C^{2} \sin^{2}{\left(\sfrac{\psi}{2} \right)}}
           {\left[C \sin{\left(\sfrac{\psi}{2} \right)} - Y{\left(t \right)}\right]^{2}
           + \left[C \cos{\left(\sfrac{\psi}{2} \right)} + X{\left(t \right)}\right]^{2}}
           \label{eq:xi_eq}\\
  \Upsilon(t) &= \frac{2 C \left[C \cos{\left(\sfrac{\psi}{2} \right)}
                + X{\left(t \right)}\right] \sin{\left(\sfrac{\psi}{2} \right)}}
                {\left[C \sin{\left(\sfrac{\psi}{2} \right)} - Y{\left(t \right)}\right]^{2}
                +\left[C \cos{\left(\sfrac{\psi}{2} \right)} + X{\left(t \right)}\right]^{2}},
                \label{eq:ups_eq}
\end{align}
where $\Xi(t)$ and $\Upsilon(t)$ are the in-phase and quadrature
components. As visualized in Fig.~\ref{fig:schematic}(b), the
instantaneous IPDs and IARs can be calculated as
$\Delta\Phi(t) = \arctantwo\left(\Upsilon(t),~\Xi(t)\right)$ and
$R(t) = \sqrt{\Upsilon(t)^{2}+\Xi(t)^{2}}$. Here, $\arctantwo$ is the
two-argument arctangent which returns the angle in the euclidean
plane.

Both Random Processes $R$ and $\Delta\Phi$ are functions of $X$ and
$Y$ which are uncorrelated Gaussian noise processes with the variance
$\sigma^{2}$. The joint PDF $f_{X, Y}(x, y)$ of $X$ and $Y$ is thus
that of a bivariate Gaussian distribution:
\begin{align}
  f_{X, Y}(x, y) = \frac{1}{2 \pi \sigma^2} \T{e}^{-\frac{1}{2\sigma^2}\left(x^2+y^2\right)}, \nonumber\\
  1 = \iint\limits^{\infty}_{-\infty}\frac{1}{2 \pi \sigma^2} \T{e}^{-\frac{1}{2\sigma^2}\left(x^2+y^2\right)} \ddiff{x}{y}\label{eq:norm_pdf},
\end{align}
to derive Eq.~\eqref{eq:norm_pdf} as a function of $\xi$ and $\upsilon$
which are instances of the processes $\Xi$ and $\Upsilon$,
Eqs.~\eqref{eq:xi_eq} and \eqref{eq:ups_eq} are rearranged to gain $x$ and
$y$ as functions of $\xi$ and $\upsilon$:
\begin{align}
  x(\xi, \upsilon) &=  C \left[
                     \frac{2 \upsilon \sin{\left(\sfrac{\psi}{2} \right)}}
                     {\upsilon^{2} + \left(\xi - 1\right)^{2}}
                     - \cos{\left(\sfrac{\psi}{2} \right)}
                     \right], \nonumber\\
  y(\xi, \upsilon) &= \frac{C \left(\upsilon^{2} + \xi^{2} - 1\right)
                     \sin{\left(\sfrac{\psi}{2} \right)}}
                     {\upsilon^{2} + \xi^{2} - 2 \xi + 1} \label{eq:xy}
\end{align}
Furthermore, by  using the Jacobian determinant $\left|J(x, y)\right|$:
\begin{align}
  \ddiff{x}{y} =\left|J(x, y)\right|\ddiff{\xi}{\upsilon}
  = \frac{4 C^{2} \sin^{2}{\left(\sfrac{\psi}{2} \right)}}
  {\left[\upsilon^{2} + \left(\xi - 1\right)^{2}\right]^{2}}
  \ddiff{\xi}{\upsilon}. \label{eq:dxdy}
\end{align}
Applying the transformations in Eqs.~\eqref{eq:xy} and \eqref{eq:dxdy}
to change the variables of Eq.~\eqref{eq:norm_pdf} results in:
\begin{align}
  1 &=
  \frac{2C^2\sin^2\left(\sfrac{\psi}{2}\right)}{\pi\sigma^2}\nonumber\\
  &\times \iint\limits^{\infty}_{-\infty}
  \frac{\T{e}^{- \frac{C^{2} \left[\upsilon^{2} - 2 \upsilon \sin{\left(\psi \right)} + \xi^{2} - 2 \xi \cos{\left(\psi \right)} + 1\right]}{2 \sigma^{2} \left[\upsilon^{2} + \left(\xi - 1\right)^{2}\right]}}}
  {\left[\upsilon^{2} + \left(\xi - 1\right)^{2}\right]^{2}}
  \ddiff{\xi}{\upsilon}.\label{eq:xiups_pdf}
\end{align}
To derive the joint PDF $f_{R, \Delta\Phi}(r, \Delta\varphi)$ which
generates the IARs $r$ and IPDs
$\Delta\varphi$. Eq.~\eqref{eq:xiups_pdf} is transformed from
rectangular to polar coordinates by using the transforms:
$\xi=r\cos{\ipd}$, $\upsilon=r\sin{\ipd}$,
$\ddiff{\xi}{\upsilon}=r\,\ddiff{r}{\ipd}$ resulting in:
\begin{align}
f_{R, \Delta\Phi}(r, \ipd) = \frac{C^22r\sin^2\left(\sfrac{\psi}{2}\right)}
  {\sigma^2\pi h(0)^{2}}\T{e}^{-\frac{C^2h(\psi)}{\sigma^2 2 h(0)}} \label{eq:ipd_ild_pdf}
\end{align}
where $h(\psi) = r^2 - 2r\cos(\ipd-\psi)+1$ and
$r\in[0, \infty], ~ \ipd\in[-\pi, \pi]$

\begin{figure*}[t]
  \centerline{\includegraphics{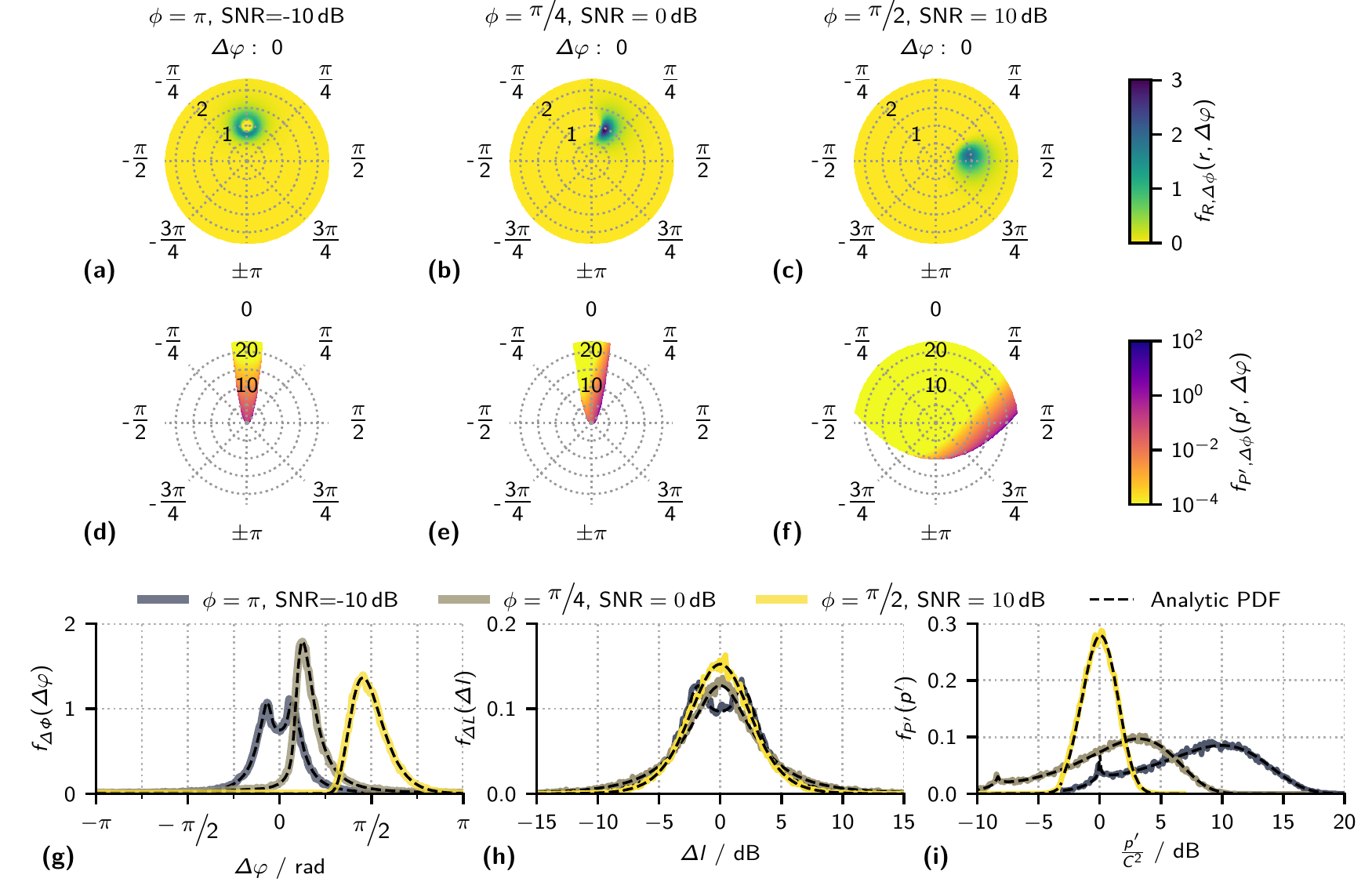}}
  \caption{%
    \textbf{(a-c)} Some examples of the joint PDF $f_{R, \Delta\Phi}$
    given in Eq.~\eqref{eq:ipd_ild_pdf}. Angles in the polar plot are
    the IPDs $\ipd$ while the radial variable is $r$, brightness/colors indicate the
    probability density.
    \textbf{(d-f)} Joint PDF $f_{P',\Delta\Phi}$ for the same
    parameters as in (a-c). As in the first row of plots, angles indicate the IPD $\ipd$ and brightness/colors the probability
    density. The radial variable however is $P'$. PDF were
    calculated for a noise variance of $\sigma^{2}=1$. A logarithmically-scaled colormap was used due to the
    large dynamic range of the PDF. White areas indicate undefined
    combinations of $P'$ and $\ipd$
    \textbf{(d, e)} Marginal PDF for $\Delta\Phi$ and $\Delta L$ and
    $P'$ compared to PDF that were estimated from the matching signal
    waveform.}
  \label{fig:verification}
\end{figure*}

This equation can be interpreted as the distribution of all possible
values of the interaural baseband $z_{1}=r\T{e}^{i \Delta\varphi}$ and
thus the distribution of all possible combinations of IPDs $\ipd$ and
IARs $r$. It is also apparent from Eq.~\eqref{eq:ipd_ild_pdf} that
equal ratios of $\sfrac{C^{2}}{\sigma^{2}}$ result in the same PDF so
that PDFs will be referenced using the signal to noise ratio
$SNR=\sfrac{C^{2}}{2\sigma^{2}}$ instead of $\sigma^{2}$ and $C$. Some
examples of these functions are shown in
Fig.~\ref{fig:verification}(a-c). Deriving the joint PDF of $\ipd$ and
ILD $\ild$ instead of IAR $r$ is easily done by using transforms
$a=10^{\sfrac{\ild}{20}}$ and $dr = \sfrac{a}{20} \ln(10)d\ild$.

To derive the joint PDF of $\Delta\Phi$ and $P'$ The process detailed
above is repeated based on the interaural baseband $Z_{2}$ as defined
in Eq.~\eqref{eq:ia_eq2} resulting in the PDF:

\begin{align}
  f_{P', \Delta\Phi}(p', \ipd) &= \frac{e^{-\frac{C^2}{2\sigma^2}
                                 -\frac{p'\left[\cos(\ipd) - \cos(\ipd - \psi)\right]}
                                 {2\sigma^2\left[\cos(\psi)-1\right]}}p'}
                                 {2\pi\sigma^2\sqrt{g}}\label{eq:pow_ipd_pdf}
\end{align}
where $g$ is given by:
\begin{align}
  g = &2 C^{2}\sin^{2}{\left(\sfrac{\psi}{2} \right)}
        \left[
        2 p' \cos{\left(\ipd \right)}
        - C^{2} \left(\cos{\left(\psi \right)} - 1\right)
        \right]
        \nonumber\\
      &- p'^{2} \sin^{2}{\left(\ipd \right)}.
\end{align}
and the range of values is defined by:
\begin{align}
  p'&\in[0, \hat{p'}(\ipd)], ~ \ipd\in[-\hat{\ipd}(p'), +\hat{\ipd}(p')]
\end{align}
where

\begin{align}
  \hat{p'}(\ipd) &= C^{2}\frac{\cos(\psi)-1}{\cos{(\ipd)}-1}.\label{eq:p_limit}
\end{align}
The function $\hat{\ipd}(p')$ can be gained by solving
Eq.~\eqref{eq:p_limit} for $\ipd$.

Similar to Eq.~\eqref{eq:ipd_ild_pdf} which defined the distribution
of all possible values of $\ipd$ and $r$, this function can be
interpreted as the distribution of all possible combinations of $\ipd$
and $p'$. The range of these combinations, however, is limited by
Eq.~\eqref{eq:p_limit} so that large areas of the exemplary PDFs shown
Fig.~\ref{fig:verification}(d--f) are undefined. This limitation will
be treated further in the discussion.

The marginal PDFs of the IAR $R$, the IPD $\Delta\Phi$ and the product
of the left and right stimulus envelope $P'$ can be calculated from
the two joint PDFs defined in Eq.~\eqref{eq:ipd_ild_pdf} and
Eq.~\eqref{eq:pow_ipd_pdf} by integrating over the other variable.
\begin{align}
  f_{\Delta\Phi}(\ipd)&=\int_{0}^{\infty}f_{R, \Delta\Phi}(r, \ipd)\diff{r}\nonumber\\
  &=\int_{0}^{\hat{p'}(\ipd)}f_{P', \Delta\Phi}(p', \ipd) \diff{p'}\label{eq:pdf_ipd}\\
  f_{R}(r)&=\int_{-\pi}^{\pi}f_{R, \Delta\Phi}(r, \ipd)\diff{\ipd} \label{eq:pdf_iar}\\
  f_{P'}(p')&=\int_{-\hat{\ipd}(p')}^{\hat{\ipd}(p')}f_{P', \Delta\Phi}(p', \ipd) \diff{\ipd}\label{eq:pdf_pow}
\end{align}
The marginal PDF for ILDs can be derived by using the transform
$r=10^{\sfrac{\ild}{20}}$:
\begin{align}
  f_{\Delta L}(\ild)&=\frac{10^{\sfrac{\ild}{20}}\log(10)}{20}\int_{-\pi}^{\pi}f_{R, \Delta\Phi}(10^{\sfrac{\ild}{20}}, \ipd)\diff{\ipd} \label{eq:pdf_ild}
\end{align}
As previously discussed, the PDFs of $\ipd$ as well as $\ild$ (and
thus $r$) only depend on the SNR and not on the absolute stimulus
power. $P'$ however, is the product of the left and right stimulus
envelope and must thus also depend on stimulus power. For this reason,
PDFs for $P'$ will always be shown normalized by $C^{2}$ so that PDFs
only depend on the SNR and are independent of overall stimulus power.

No closed-form solution for Eq.~\eqref{eq:pdf_ipd}--\eqref{eq:pdf_ild}
could be found so that numeric integration was used to evaluate them
(QUADPACK algorithms QAGS/QAGI
\cite{Piessens1983}). Figs.~\ref{fig:verification}(g)--(i) show some
examples of the PDF of $\Delta\Phi$, $\Delta L$, $P'$ and verifies the
results by comparing Eq.\eqref{eq:pdf_ipd}--\eqref{eq:pdf_iar}
to PDFs that were numerically estimated from signal waveforms.

\section{Discussion}
All PDFs derived above show discontinuities for $\ipd\in\{0, \pm\pi\}$
for which the probability densities approach zero. Or in other words,
a \NphiSphi{0}{\psi} stimulus will never contain IPDs that are exactly
zero or $\pi$. Both discontinuities can be understood when keeping in
mind that the IPD is defined by
$\ipd = \arctantwo\left(\upsilon,~\xi\right)$. Which can %
only result in a value of $0$ or $\pm\pi$ if $\upsilon=0$. This is
only the case when $x=-C\cos\left(\sfrac{\psi}{2}\right)$. As the
probability of $x$ to take this exact value approaches zero, the joint
PDFs will also approach zero. For further discussion of the PDFs
however, this discontinuity will not be shown explicitly in plots
as it's implication in practice is limited.

\begin{figure*}[t]
  \includegraphics{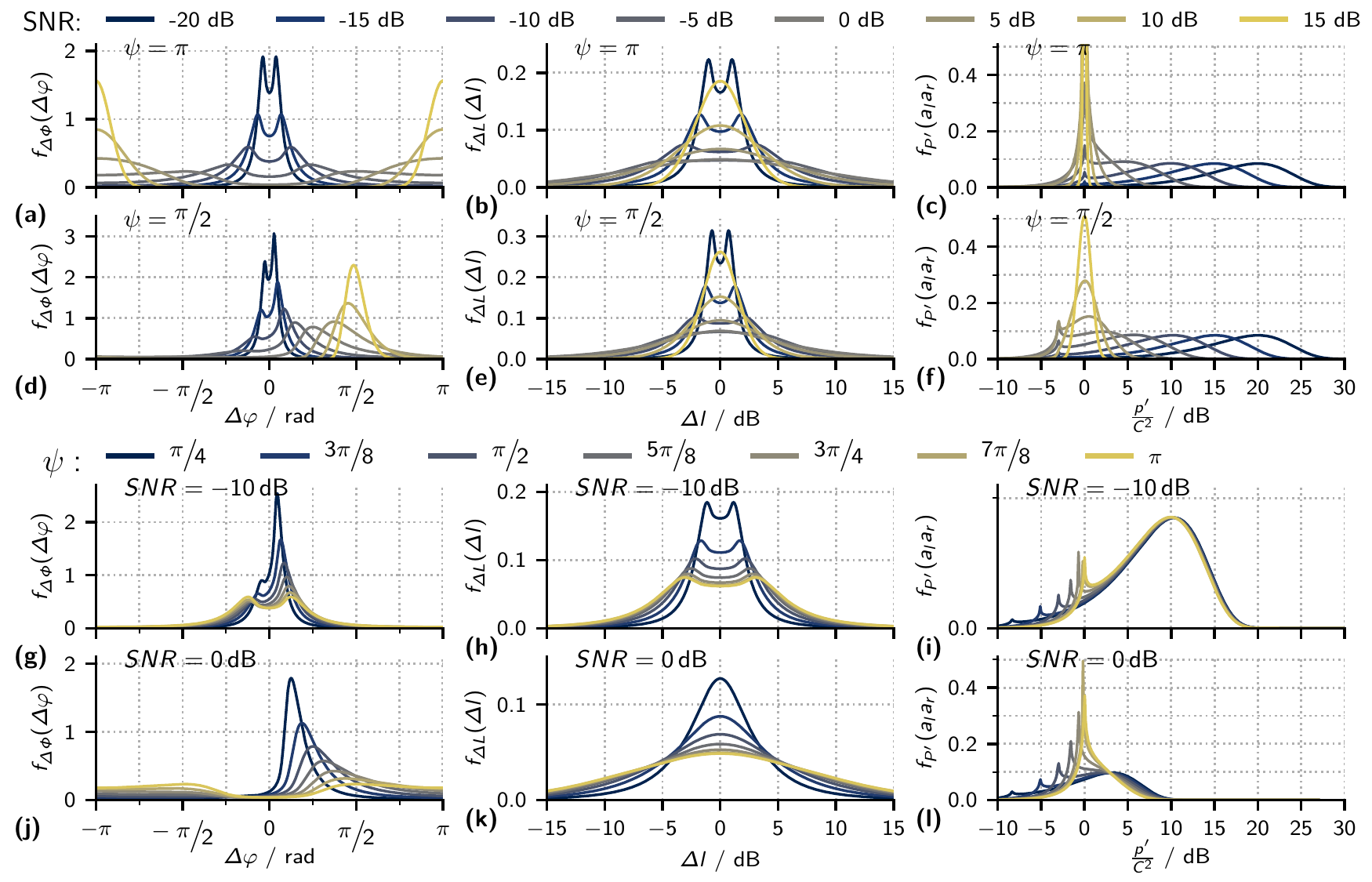}
  \caption{Exemplary marginal PDFs \textbf{(a)--(f)} Calculated for two fixed
    signal phases $\psi$  at different SNRs (color/brightness-coded)
    and \textbf{(g)--(l)} calculated at two fixed SNR for different
    signal phases (color/brightness coded).}\label{fig:example}
\end{figure*}

Figure~\ref{fig:example}(a) and (d) show examples of the marginal IPD
PDFs for $\psi=\pi$ and $\psi=\sfrac{\pi}{2}$ while varying the
SNR. The instantaneous IPD $\ipd$ can be interpreted as a result of
the mixture of zero IPD due to the diotic noise and the IPD $\psi$ of
the tone. The weighting of the two IPDs is determined by the
instantaneous power of the noise relative to the power of the
tone. Thus, at large negative SNRs where the stimulus is dominated by
noise, IPD PDFs show a mean value close to zero and only little
variance. With increasing SNR, the IPDs are increasingly influenced by
the tone-IPD so that the distributions mean moves towards $\psi$ and
variance increases. At larger positive SNRs, where the noise power is
small compared to the tone, the IPDs are dominated by the tone-IPD
$\psi$ so that the variance decreases again. In the two extreme cases
where the SNR would either be $-\infty$ or $+\infty$, the signal
consists of only the noise or the tone so that neither IPD nor ILD
fluctuates - both PDFs are then $\delta$-distributions. For the IPD,
this distribution is either be located at zero (SNR=$-\infty$) or at
$\psi$ (SNR=$+\infty$) while the ILD distribution is always centered
at \SI{0}{\decibel}. ILD PDFs for the same parameters as used for the
IPD PDFs in Fig.~\ref{fig:example}(a),(d) are shown in the panels (b)
and (e) of the same Figure. Instantanious ILDs $\Delta l$, are a
direct result of the relative energy of the instantanious noise and
the tone. As a result, ILD PDFs exhibit the same change of variance as
discussed for the IPDs, low variance at both high or low SNR where the
stimulus is either dominated by the tone or noise and an increase of
variance at intermediate SNRs. Figure~\ref{fig:example}(c) and (f)
show distributions for the remaining parameter $P'$ plotted in decibel
relative to the squared amplitude of the tone. For large SNRs, the
signal is dominated by the tone, $\sfrac{p'}{C^{2}}$ is thus narrowly
distributed around \SI{0}{\decibel}. With decreasing SNR, the noise
power increases relative to $C^{2}$ so that the peak of the
distribution shifts towards larger values of $\sfrac{p'}{C^{2}}$ with
the overall shape of the distribution remaining largely unchanged.

Figures~\ref{fig:example}(g)--(l) additionally show IPD, ILD and $P'$
PDFs for cases where the SNR was fixed while varying $\psi$. From the
vector summation shown in Fig.~\ref{fig:schematic}(b), it is intuitive
that, at the same tone amplitude $C$, a smaller value of $\psi$ also
results in smaller IPDs. As a direct consequence, IPD and ILD PDFs
also show less variance for smaller values of $\psi$. The PDFs for P',
however, are largely uninfluenced by $\psi$ - with the notable
exception of a sharp peak located at
$\sfrac{p'}{C^{2}}=\sin^{2}(\sfrac{\psi}{2})$. This peak is a
consequence of Eq.~\eqref{eq:p_limit} which limit's the possible
combinations of IPDs and $P'$.

\begin{figure}[h]
  \centering
  \includegraphics{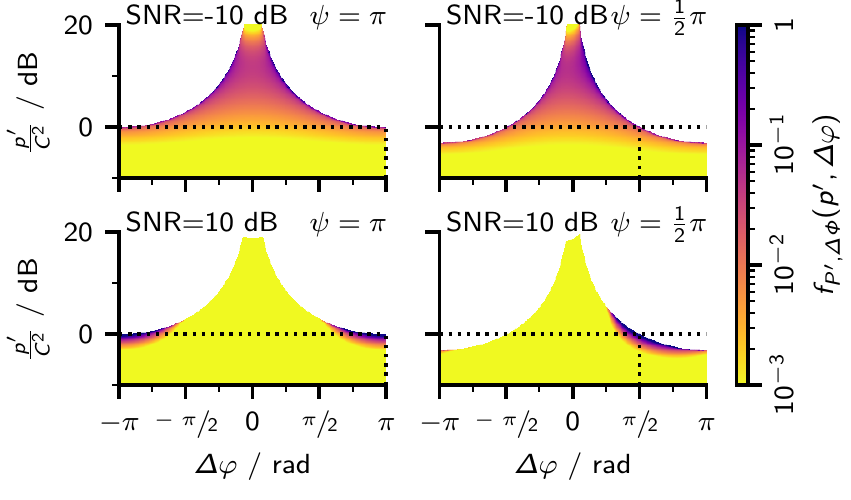}
  \caption{Joint probability functions of $P'$ and IPD as defined in
    Eq.~\eqref{eq:pow_ipd_pdf}. The panels show functions for
    different SNRs and Tone-IPDs $\psi$. The horizontal dashed black
    lines indicate the location where $p'=C^{2}$, the vertical black
    lines indicate where $\ipd=\psi$. Note that the color-map is
    logarithmically-scaled and that the changes in the scale were
    limited to values between 1 and below $10^{-3}$.}\label{fig4}
\end{figure}

Figure~\ref{fig4} shows some exemplary joint PDFs of IPD and $P'$ with
the undefined region shown in white. It is notable that the
probabilities are heavily clustered close to the limit defined by
Eq.~\eqref{eq:p_limit}. The density in these functions decreases so
quickly that a logarithmically-scaled color map had to be chosen in
order to visualize the function. The limit itself is a consequence of
the vector summation visualized in Fig.~\ref{fig:schematic}(b) and is
best discussed for $\ipd=\psi$. This is obviously true for any point
in time where the noise energy is zero so that the signal only
contains the tone and $p'=C^{2}$. Using geometry it can then be
shown that $p'<C^{2}$ for any other case where $\ipd=\psi$. Thus
follows that $\hat{p}'(\ipd=\pm\psi)=C^{2}$, a relation that is also
visualized in form of the dotted black lines in Fig.~\ref{fig4}. The
large probability density close to the limit defined by
Eq.~\eqref{eq:p_limit} also explains the sharp peak in the marginal
PDF of $P'$ as visible in Fig.~\ref{fig:example}.  From
Eq.~\eqref{eq:p_limit} follows that
$\hat{p}'(\ipd=\pm\pi)=C^{2}\sin^{2}(\sfrac{\psi}{2})$ thus resulting
in a peak around this location.

All PDFs derived in this study are independent of the noise
spectrum. The spectral properties and especially the bandwidth does,
however, influence the frequency of IPD, ILD and $P'$
fluctuations. Larger bandwidths result in faster
fluctuations. Further, the tone does not need to be spectrally
centered in the noise. It does not even have to be within the noise
spectrum. With auditory processing, especially peripheral filtering,
the spectrum is of course going to influence the SNR at the level of
binaural interaction and thus the PDFs of the encoded binaural cues.

While all PDFs were derived for the diotic noise case
\NphiSphi{0}{\psi}, they can easily be generalized to cases where an
additional phase delay $\psi_{2}$ is applied to the whole
stimulus. Such a signal could then be referred to as
$(\NphiSphi{0}{\psi})_{\psi_{2}}$ and would result in the same IPD
distributions as in the $\NphiSphi{0}{\psi}$ case but shifted by
$\psi_{2}$ with ILD and $P'$ distributions remaining unchanged.

\section{Summary}
Goal of this study was to derive the joint PDF for ILDs (IARs) and
IPDs as well as IPDs and $P'$. The two functions are given by the
Eqs.~\eqref{eq:ipd_ild_pdf} and \eqref{eq:pow_ipd_pdf}. They are a key
component for understanding how the SNR and $\psi$ influence the
magnitude of binaural unmasking.

\section*{Acknowledgments}
This work was supported by the European Research Council (ERC) under the European Union’s Horizon 2020 Research and Innovation Programme grant agreement No. 716800 (ERC Starting Grant to M.D.)

\bibliographystyle{bibstyle-new}
\bibliography{library}
\end{document}